\documentclass[sigconf]{acmart}

\acmConference[EASE 2027]{The 31st International Conference on Evaluation and Assessment in Software Engineering}{15–18 June, 2027}{Hanoi, Vietnam}
\AtBeginDocument{%
  }

\usepackage{amsmath}
\usepackage{enumitem}
\usepackage{tcolorbox}
\usepackage[htt]{hyphenat}
\usepackage{multirow}
\usepackage{hyperref}
\usepackage{xcolor}

\setcopyright{acmlicensed}
\copyrightyear{2018}
\acmYear{2018}
\acmDOI{XXXXXXX.XXXXXXX}
\acmConference[Conference acronym 'XX]{Make sure to enter the correct
  conference title from your rights confirmation email}{June 03--05,
  2018}{Woodstock, NY}
\acmISBN{978-1-4503-XXXX-X/2018/06}

\begin{document}

\title{Will It Survive? Deciphering the Fate of AI Agent-Authored Code in Open Source}

\author{Musfiqur Rahman}
\affiliation{%
    \institution{Concordia University}
    \city{Montr\'eal}
    \country{Canada}
}
\email{musfiqur.rahman@mail.concordia.ca}

\author{Mahi Begoug}
\affiliation{%
    \institution{Concordia University}
    \city{Montr\'eal}
    \country{Canada}
}
\email{mahi.begoug@concordia.ca}

\author{Emad Shihab}
\affiliation{%
    \institution{Concordia University}
    \city{Montr\'eal}
    \country{Canada}
}
\email{emad.shihab@concordia.ca}

\renewcommand{\shortauthors}{Rahman et al.}

\begin{abstract}
AI coding agents such as GitHub Copilot, OpenAI Codex and Devin now open pull requests (PRs) that are merged into open-source projects. Industry reports suggest that code written with AI assistance is revised or reverted soon after it is written, which raises the concern that agent code is ``disposable''. However, very little is known about what happens to the code of coding agents after it is merged. Understanding how long this code survives, and why it is changed, can help maintainers decide how much review such code needs. To address this gap, we performed a survival analysis of the 648,959 lines that 2,794 merged PRs added outside vendored and generated paths in 215 open-source repositories from the AIDev dataset, comparing agent-attributed and human code within the same repositories. We find that agent-attributed code is not modified sooner than human code. When each PR is weighted equally, agent-attributed lines are modified later (HR = 0.83, 95\% CI [0.75, 0.93]). This difference comes from the supervised assistants, mainly Copilot and Codex (HR = 0.76). Code from Devin does not differ from human code over the whole period (HR = 0.95), but among PRs merged after mid-May 2025, code from both the supervised assistants and Devin is modified later. However, without PR weighting, the difference is not statistically significant. When agent-attributed code is modified, the intent of the change is similar to that of human code (Cram\'er's V = 0.080). Finally, a file's size and recent changes predict whether agent-attributed code will be modified better than its code tokens do (AUC-ROC 0.78 vs.\ 0.70). Process features predict when it will be modified only weakly (Macro F1 of 0.31 against 0.25 for a random guess). Our findings suggest that agent code is not disposable, but whether it survives longer than human code depends on the agent, on when the code was merged and on how the comparison is computed.
\end{abstract}




\maketitle

\section{Introduction}
\label{sec:introduction}

Nowadays, AI coding agents such as GitHub Copilot, OpenAI Codex, Cursor, Claude Code and Devin write code in real-world software projects. They open pull requests (PRs) that are reviewed and merged like any other contribution. For example, the AIDev dataset contains 932,791 PRs opened by five agents on GitHub~\cite{li2026aidev}, and 76\% of the respondents of the 2024 Stack Overflow Developer Survey use or plan to use AI tools in their development process~\cite{stackoverflow2024Stack}.

However, these tools are evaluated almost exclusively at generation time, with metrics such as Pass@k, CodeBLEU or compilation rates~\cite{chen2021evaluating,ren2020codebleu}. These metrics tell us whether an agent can produce correct code, but not what happens to that code after it is merged. This is important because maintenance accounts for a large share of the total cost of software~\cite{boehm1984software}. Therefore, if agent code passes review but is modified or rewritten soon after the merge, the time saved while writing it may be spent again on maintaining it.

GitClear analyzed about 153 million changed lines from 2020 to 2023 and reported that code \textit{churn}, the share of lines that are reverted or updated less than two weeks after they are written, rises as AI coding assistants spread~\cite{gitclearCodingCopilot}. If this also holds for code written by coding agents, agent code is effectively ``disposable'': merged quickly, but modified or deleted soon after. In this study, we refer to this concern as the \textit{disposable code hypothesis}. However, very little is known about what happens to the code of coding agents after it is merged into real projects.

In this study, the term \textit{agent-attributed code} refers to the code added by a merged PR that the AIDev dataset attributes to a coding agent, and \textit{human code} refers to the code added by a merged PR that AIDev lists as written by a human. We use ``attributed'' because a PR-level signal shows why a PR was collected, but not who wrote each line, and we audit this attribution on a sample of PRs (Section~\ref{sec:attribution}).

To test the disposable code hypothesis, we use survival analysis, which models the time until an event while handling units whose event has not yet happened~\cite{kleinbaum1996survival}. Survival analysis has been used in software engineering for a long time (Section~\ref{sec:related}), for example to estimate how long lines of code live~\cite{spinellis2021software}, and recent studies apply it to code from coding agents~\cite{xia2026these}. We follow the lines added by 2,802 merged PRs, 1,470 agent-attributed and 1,332 human, in 215 open-source repositories from AIDev, from the merge until the line is first modified or until December 31, 2025. Agent and human PRs are spread unevenly across repositories, so we compare agent-attributed and human code within the same repository. Furthermore, a few very large PRs contribute most of the lines, so our primary analysis gives each PR the same weight. We also separate the \textit{supervised assistants} (Copilot, Codex, Cursor and Claude Code), which work on tasks that a developer starts and reviews, from Devin, an autonomous software engineer. We ask three research questions:

\textbf{RQ1 (Survival): Is agent-attributed code modified sooner than human code?}
Agent-attributed code is not modified sooner than human code in the same repositories in any of the analysis variants we examined (e.g., with or without weighting, at the line, hunk or file level). In our primary analysis, which gives every PR the same weight regardless of how many lines it adds, lines from supervised assistants, mainly Copilot and Codex, are modified later than human lines (hazard ratio, HR = 0.76 [0.65, 0.87], where an HR below 1 means that agent-attributed code is modified later). Devin code does not differ from human code in the full cohort (HR = 0.95), although among PRs merged after mid-May 2025, when Codex was released, Devin code is also modified later. When every line counts equally instead, so that a few very large PRs dominate, neither difference is significant.

\textbf{RQ2 (Intent): When agent-attributed code is modified, what is the intent of the modification?}
The commits that modify agent-attributed code have almost the same intent profile as those that modify human code (Cram\'er's V = 0.080). The shares of the maintenance categories of Swanson~\cite{swanson1976dimensions}, perfective (improvements and new features, 38.3\% vs.\ 35.4\%), corrective (bug fixes, 35.6\% vs.\ 32.5\%), adaptive (changes to the environment, 11.7\% vs.\ 12.7\%) and preventive (testing and security, 3.9\% vs.\ 3.5\%), are close, and only unclassifiable commits are less common for agent-attributed code (10.6\% vs.\ 15.9\%).

\textbf{RQ3 (Forecasting): Can we predict at merge time whether, and when, agent-attributed code will be modified?}
Code tokens predict whether an agent-attributed file will be modified with modest accuracy (AUC-ROC 0.70), but a baseline on the size and recent changes of the file predicts it better (AUC-ROC 0.78), and the tokens that the model relies on do not point to the lines that are later modified. Process features, such as the recent changes to the file and its age, predict in which of four time bins (from within a day to more than a month) the code will be modified only weakly (Macro F1 0.31 against 0.25 for chance).

\smallskip
\noindent In summary, this paper makes the following contributions:
\begin{itemize}[leftmargin=*,nosep]
    \item We find that agent-attributed code is not modified sooner than human code in the same repositories, and that the lower rate of modification of supervised-assistant code holds only when every PR gets the same weight.
    \item We find that the commits that modify agent-attributed code have almost the same intent as those that modify human code, and that matching keywords inside other words (e.g., \textit{auth} inside the \texttt{Co-authored-by} trailer) creates a spurious difference.
    \item We find that the size and recent history of a file predict whether agent-attributed code will be modified better than its code tokens do, and that the code tokens the model relies on do not point to the lines that are later modified. Process features predict the timing of a modification only weakly.
    \item We release our data, attribution audit and scripts to support replication (see the \nameref{sec:replication} section).
\end{itemize}

\section{Methodology}
\label{sec:methodology}

\subsection{Dataset}

\subsubsection{Source Dataset}
We use the AIDev dataset~\cite{li2026aidev}, which collects PRs opened by five AI coding agents in public GitHub repositories up to August 1, 2025. Of its 932,791 agent PRs, 814,522 come from OpenAI Codex, 50,447 from GitHub Copilot, 32,941 from Cursor, 29,744 from Devin and 5,137 from Claude Code, in 116,211 repositories. In the AIDev data, agent PRs carry PR-level signals of the agent, such as the prefix of the head branch (\texttt{codex/}, \texttt{copilot/} and \texttt{cursor/}), the bot account that opens the PR (\texttt{devin-ai-integration[bot]} and \texttt{Copilot}) or a \texttt{Co-Authored-By: Claude} line in the PR description. AIDev also contains 6,618 human PRs in 818 repositories, created between January and June 2025. We keep only repositories that contain both agent and human PRs, so that agent-attributed and human code can be compared under the same project conditions, such as coding standards and review practices.

\subsubsection{Repository Filtering}
We apply the filtering criteria of Xiao et al.~\cite{xiao2026self}, which select engineered software projects~\cite{munaiah2017curating} (Table~\ref{tab:funnel}). We (1) keep repositories with both agent and human PRs, (2) remove repositories without a software license, (3) remove archived repositories, repositories without releases or tags and repositories with fewer than two contributors, (4) remove, per language, repositories in the bottom quartile of PR count, open issue count or size, and (5) remove repositories whose share of code lines among code and comment lines lies outside the central 97\% of that share among repositories of the same language. We keep the 97\% interval of Xiao et al. so that our cohort follows a published definition of engineered projects.

\begin{table}[t]
\centering
\caption{Repository filtering funnel and mined cohort}
\label{tab:funnel}
\resizebox{\columnwidth}{!}{
\begin{tabular}{lrrr}
\toprule
\textbf{Step} & \textbf{Repos} & \textbf{Agent PRs} & \textbf{Human PRs} \\
\midrule
AIDev, all repositories & 116,211 & 932,791 & 6,618 \\
(1) Repositories with agent and human PRs & 816 & 9,797 & 6,598 \\
(2) Software license & 666 & 7,625 & 5,054 \\
(3) Not archived, has release or tag, $\geq$2 contributors & 566 & 6,891 & 4,559 \\
(4) Bottom-quartile removal per language & 266 & 4,649 & 3,196 \\
(5) Code ratio within the 97\% interval & 256 & 4,563 & 3,132 \\
\midrule
Repositories with at least one merged PR & 246 & 2,302 & 2,386 \\
Repository history available (Section~\ref{sec:cohort}) & 244 & 2,266 & 2,351 \\
Not listed as both agent and human PR & 244 & 2,261 & 2,346 \\
Merge commit located on the default branch & 241 & 2,092 & 2,156 \\
Adds at least one tracked source line (mined cohort) & 215 & 1,470 & 1,332 \\
\bottomrule
\end{tabular}}
\parbox{\linewidth}{\footnotesize \vspace{0.3em} AIDev's 6,618 human PRs include 20 in two repositories without agent PRs. Step 2 also removes three repositories whose GitHub metadata could not be retrieved, and a minimum of 500 stars, applied before step 2, removes no repository. Steps 1 to 5 count all PRs, merged or not. The rows below the line count merged PRs only. A tracked source file is a file with a source-code extension, excluding configuration and documentation files.}
\end{table}

\subsubsection{Final Cohort}
\label{sec:cohort}
Of the 256 filtered repositories, 246 have at least one merged PR. We mine every merged PR whose merge commit we can locate on the default branch and that adds at least one line to a tracked source file. Table~\ref{tab:funnel} shows the attrition. Two repositories cannot be mined, because \texttt{dust-tt/dust} is no longer public and \texttt{elizaos/eliza} rewrote its history after our observation window. We also exclude five PRs that AIDev lists as both agent and human PRs, we could not locate the merge commit of 359 PRs, and 1,446 PRs change no tracked source file (e.g., documentation only). The mined cohort contains \textbf{2,802 merged PRs in 215 repositories}: 1,470 agent-attributed PRs (492 from Copilot, 491 from Codex, 412 from Devin, 44 from Cursor and 31 from Claude Code) and 1,332 human PRs. Eight PRs add only vendored or generated lines, that is, third-party code or build output that is committed to the repository but not written by hand (Section~\ref{sec:vendored}), so our main analysis uses 2,794 PRs.

\subsubsection{Agent Tiers}
\label{sec:tiers}
The agents differ in how a developer works with them. Copilot's coding agent works on an issue that a developer assigns to it and reviews~\cite{copilotAgentDocs}, Codex works on a task that a developer starts and reviews before opening the PR~\cite{codexDocs}, and Cursor and Claude Code run in the developer's editor or terminal, or as background agents that the developer starts~\cite{cursorAgentDocs,claudeCodeDocs}. We call these four agents \textit{supervised assistants}. In contrast, Devin is offered primarily as an autonomous software engineer that takes tasks from a team~\cite{devinDocs}, and its PRs in AIDev are opened by its own bot account (\texttt{devin-ai-integration[bot]}). We refer to these two groups, the supervised assistants and Devin, as agent tiers. We describe the primary workflow of each product in 2025, although some products now also accept tasks from other tools. We acknowledge that the tiers describe how the agents are offered, not how each PR was supervised. For this reason, we also report RQ1 for each agent.

\subsubsection{Attribution Validation}
\label{sec:attribution}
AIDev's signals do not prove that the agent wrote the code of a PR. For example, a developer may push commits to an agent PR. Therefore, we audited a random sample of agent PRs. From the 2,092 merged agent PRs whose merge commit we located (Table~\ref{tab:funnel}), we drew 325 PRs, which is the sample size for a proportion at 95\% confidence and a $\pm$5\% margin with the finite-population correction~\cite{cochran1977sampling}, stratified proportionally by agent. For each PR, the first author inspected the commits that change source files and labeled the PR \textit{agent} (every substantive code change comes from the agent), \textit{mixed} (the agent and a human both made substantive changes), \textit{human} (all substantive changes come from a human) or \textit{unclear} (the history cannot be inspected or its signals conflict), following a written guide that is part of our replication package. A change is non-substantive only if it belongs to one of the non-essential change categories of Kawrykow and Robillard~\cite{kawrykow2011non}, such as whitespace changes or reference updates caused by a rename. A second rater independently labeled 100 of the 325 PRs (31\%) with the same guide, as Ehsani et al.~\cite{ehsani2026aicodingagentsfail} did for agent PRs. This exceeds the 10 to 25\% that is typical for estimating agreement~\cite{oconnor2020intercoder} and the 62 PRs needed to distinguish $\kappa = 0.8$ from 0.5 with four labels~\cite{bujang2017guidelines}. The raters agree almost perfectly (Cohen's $\kappa$ = 0.85)~\cite{cohen1960coefficient,landis1977measurement} and resolved the disagreements by discussion. We report 95\% Wilson score intervals~\cite{brown2001interval}.

Of the 325 audited PRs, 218 (67.1\% [61.8, 72.0]) are labeled agent, 71 (21.8\%) mixed, 20 (6.2\%) human and 16 (4.9\%) unclear, so the agent made substantive changes in 88.9\% of them. Not every audited PR is in our mined cohort, because many change no tracked source file. Within the cohort, the agent made substantive changes in 87.3\% [82.3, 91.1] of the audited PRs (98.4\% of the 64 Devin PRs and 82.8\% [76.1, 87.9] of the 157 supervised-assistant PRs, so the attribution is noisier for the supervised assistants), and 16 PRs are labeled human. As a sensitivity analysis, RQ1 repeats the analysis without these 16 PRs.

\subsection{Measuring Survival}
\label{sec:survival}

We frame the modification of code as a survival problem: a unit of code survives until it is modified, and the \textit{modification event} ends its survival time. Survival analysis suits this problem because it handles right-censored units, that is, code that is still unmodified when the observation window ends~\cite{kleinbaum1996survival}.

\subsubsection{Definitions}
\begin{itemize}[leftmargin=*]
    \item \textbf{Birth:} a unit is born when its PR is merged into the default branch ($t=0$).
    \item \textbf{Modification event:} a unit is modified when a later commit on the default branch changes or deletes it ($t>0$).
    \item \textbf{Censoring:} units that are still unmodified on December 31, 2025 are right-censored. Because AIDev contains PRs up to August 1, 2025, every unit is observed for at least five months, which covers the two-week window of the churn metric that motivates the disposable code hypothesis~\cite{gitclearCodingCopilot} many times over.
\end{itemize}
\noindent A modification event is neutral. It does not imply a defect, since code is also modified to add functionality or to adapt to its environment, which RQ2 examines. Each unit is followed from its own merge date, so agent-attributed and human code merged in the same period are exposed for the same time.

\subsubsection{Units of Analysis}
\paragraph{Line (primary unit)} Every line that a PR adds to a source file. We follow all of these lines rather than a sample of them. We follow each line with \texttt{git blame --reverse} from the merge commit to the last commit on the default branch on or before December 31, 2025. The blame reports the last commit in which the line still exists, and the line is modified by the next commit that touches its file after that commit. Because each line is attributed to its own PR, a change to a human line elsewhere in the same file does not count against an agent line.
\paragraph{Hunk} A contiguous block of lines that the PR adds in one diff hunk. A hunk is modified at the first time any of its lines is modified.
\paragraph{File} A source file that the PR changes. A file is modified when any later commit touches it. A file usually mixes lines from many authors and PRs, so we report the file level only for comparison with file-level studies~\cite{sawada2026extent}.

\subsubsection{Substantial Modification}
Under the definitions above, even a one-character edit ends the survival of a unit. To separate minor edits from rewrites, we also define a \textit{substantial modification} event for hunks and files: a unit is substantially modified at the first time a share $\theta$ of its added lines, that is, the lines that the PR added to it rather than all the lines of the file, has been modified or deleted. We are not aware of an established value for $\theta$, so instead of choosing one, we report the whole range from a quarter of the lines to all of them, $\theta \in \{25\%, 50\%, 75\%, 100\%\}$.

\subsubsection{Vendored and Generated Code}
\label{sec:vendored}
Repositories often commit code that is not written by hand, such as third-party libraries and build output, and regenerate it in bulk. A regenerated file shows up as a modification of all its lines, which measures how often a project rebuilds its output rather than how its code is maintained. Therefore, our primary analysis excludes lines in vendored and generated paths, which we identify by their path: directories named \texttt{vendor}, \texttt{third\_party}, \texttt{node\_modules}, \texttt{dist}, \texttt{build}, \texttt{out}, \texttt{generated} or \texttt{\_\_generated\_\_}, minified and bundled JavaScript files, and generated protocol buffer and designer files. Most of these patterns come from GitHub Linguist~\cite{githubLinguist}, which GitHub uses to exclude such files from its language statistics, and we add common build output directories. The list is not exhaustive, so we also report the analysis with all lines.

\subsubsection{Weighting}
A few very large PRs contribute a large share of all lines. For example, the largest human PR in our cohort alone adds more than 34,000 lines outside vendored and generated paths. Let us take an arbitrary example to show why this matters. Suppose that a repository has 100 human PRs. One PR adds 10,000 lines, which are reformatted a week later, and each of the other 99 PRs adds 10 lines, which are never modified. If every line has the same weight, 91\% of the human lines are modified (10,000 of 10,990), so human code looks short-lived. If every PR has the same weight, the code of only one PR in 100 (1\%) is modified, so human code looks stable. In other words, a line-weighted comparison describes a few large PRs, whereas our question concerns the code that agent and human PRs typically contribute. Therefore, our primary analysis weights each unit by the inverse of the number of units in its PR, so that every PR has the same weight (\textit{PR-weighted}). We report the unweighted analysis next to it in every table, so that the reader can see where the two differ. We call a combination of a unit of analysis, a set of paths and a weighting a \textit{design}. Our \textit{primary design} follows the lines outside vendored and generated paths with PR weighting, and we treat every other design as a sensitivity analysis (Section~\ref{sec:rq1}).

\subsubsection{Implementation Details}
We track source files by extension and exclude configuration and documentation files. We locate the merge commit of each PR on the default branch by matching GitHub's merge time against its first-parent history, which contains the commits of the default branch but not those inside merged branches, and we use the PR number in the commit message to break ties, which also covers squash and rebase merges. Table~\ref{tab:survival_stats} summarizes the units and their modification events.

\begin{table}[t]
\centering
\caption{Units of analysis and modification events (primary design). Percentages are PR-weighted, with unweighted values in parentheses.}
\label{tab:survival_stats}
\resizebox{\columnwidth}{!}{
\begin{tabular}{lrrr}
\toprule
\textbf{Metric} & \textbf{Line} & \textbf{Hunk} & \textbf{File} \\
\midrule
Units & 648,959 & 60,095 & 18,120 \\
Modified, human & 34.5\% (30.1\%) & 44.8\% (39.3\%) & 88.2\% (85.2\%) \\
Modified, supervised assistants & 24.1\% (23.3\%) & 34.5\% (30.9\%) & 81.3\% (77.1\%) \\
Modified, Devin & 38.0\% (23.6\%) & 52.5\% (51.1\%) & 91.5\% (85.6\%) \\
Median days to modification (modified units) & 40.1 (39.0) & 35.9 (40.1) & 10.0 (11.6) \\
Median follow-up of censored units (days) & 218.3 & 236.5 & 216.5 \\
\bottomrule
\end{tabular}}
\parbox{\linewidth}{\footnotesize \vspace{0.3em} The 2,794 PRs of the primary design add 648,959 lines outside vendored and generated paths. Devin's unweighted line rate is lower than its PR-weighted rate because its large PRs are modified less often than its small ones, and its largest PR adds 41.6\% of its lines.}
\end{table}

\section{RQ1: Is Agent-Attributed Code Modified Sooner Than Human Code?}
\label{sec:rq1}

\subsection{Motivation}
The disposable code hypothesis is our own, derived from GitClear's churn finding~\cite{gitclearCodingCopilot} (Section~\ref{sec:introduction}). Under this hypothesis, code from coding agents is merged quickly but modified or deleted soon afterwards. If the hypothesis holds, agent-attributed code should reach a modification event sooner than human code merged into the same repositories. Testing this is the first step to know whether code from coding agents adds to the maintenance burden of a project.

\subsection{Approach}
\paragraph{Kaplan-Meier estimation and log-rank test}
We estimate the survival function $S(t)$, the probability that a unit is still unmodified $t$ days after its merge, for each group with the Kaplan-Meier estimator~\cite{kaplan1958nonparametric}, and compare the agent-attributed and human curves with the log-rank test~\cite{mantel1966evaluation}.

\paragraph{Within-repository Cox regression}
Agent and human PRs are not spread evenly across repositories, and repositories differ in how often they change their code. Therefore, a model pooled across repositories would partly compare active and quiet projects rather than agent and human code. Our primary model is a Cox proportional hazards model~\cite{cox1972regression} stratified by repository, so that each repository has its own baseline hazard and agent-attributed code is compared only with human code in the same repository:
\begin{equation}
\begin{split}
h_r(t \mid X) = h_{0,r}(t) \cdot \exp\big(&\beta_1 \cdot \texttt{is\_agent} + \beta_2 \log(1+\texttt{added})\\
&+ \beta_3 \log(1+\texttt{files})\big),
\end{split}
\end{equation}
where $r$ is the repository, and \texttt{added} and \texttt{files} are the lines added and the source files changed by the PR. Lines from the same PR are not independent, so we use robust standard errors clustered by PR~\cite{lin1989robust}. A hazard ratio (HR) $\exp(\beta_1)$ below 1 means that agent-attributed code is modified later than human code in the same repository. A second model replaces \texttt{is\_agent} with one indicator per tier, and a third with one indicator per agent. We test the proportional hazards assumption with the Schoenfeld residual test~\cite{schoenfeld1982partial}.

\paragraph{Restricted mean survival time}
An HR is hard to read when the hazards are not proportional. Therefore, we also report the restricted mean survival time (RMST)~\cite{royston2013restricted}, the mean number of days within the first $\tau$ days that a unit stays unmodified, which needs no proportional hazards assumption. We use $\tau = 90$ days and $\tau = 150$ days, the minimum follow-up of every unit. We obtain 95\% confidence intervals for the agent-human difference in RMST and in the share of modified units from a bootstrap that resamples repositories 1,000 times.

\paragraph{Sensitivity analyses}
Our primary design rests on choices that other studies could make differently. To check that our conclusion does not depend on them, we repeat the primary analysis (i) without weighting, (ii) with vendored and generated paths, (iii) with a sampled design that follows at most 20 source files per PR and 100 random lines per file, which caps the influence of very large PRs in a second way, (iv) at the hunk and file levels, (v) with the substantial modification event, (vi) with covariates of each line's file at merge time (changes in the previous three months, age, created by the PR, test file), because the history of a file may predict when it changes, (vii) without the PRs that the audit labeled human, or human and unclear, and with only the PRs labeled agent, (viii) within terciles of repository activity (the number of PRs of the repository in AIDev), and (ix) within repository and task type, because agent and human PRs may address different tasks. For (ix), we use the conventional-commit type (e.g., \textit{feat}, \textit{fix}, \textit{refactor}) that AIDev assigns to each PR from its title and description~\cite{li2026aidev}, and pool the rare types as \textit{other}. These labels cover 2,781 of our 2,794 PRs. They describe PRs, not commits, so they cover only 8.5\% of the 5,600 commits that modify our lines, which is why RQ2 classifies commit messages instead. We also fit a Cox model pooled across repositories, adjusted for PR size, stars and contributors, with PR-clustered standard errors.

\subsection{Results}

\paragraph{Primary analysis}
Table~\ref{tab:rq1_line} reports the results. In the primary analysis, agent-attributed lines are modified later than human lines in the same repository (HR = 0.83 [0.75, 0.93], $p = 0.001$). This difference comes from the supervised assistants (HR = 0.76 [0.65, 0.87], $p < 0.001$), whereas Devin does not differ from human code (HR = 0.95 [0.82, 1.11], $p = 0.53$). In other words, the hazard of modification of a supervised-assistant line is 24\% lower than that of a human line in the same repository. By the end of the window, 28.0\% of agent-attributed lines and 34.5\% of human lines are modified (PR-weighted), a difference of $-6.5$ percentage points (pp) [$-10.7$, $-1.7$].

\begin{table}[t]
\centering
\caption{RQ1: Within-repository Cox HR [95\% CI] of agent-attributed vs.\ human code. An HR below 1 means that agent-attributed code is modified later.}
\label{tab:rq1_line}
\resizebox{\columnwidth}{!}{
\begin{tabular}{lrrrrr}
\toprule
\textbf{Design} & \textbf{Units} & \textbf{Mod.\ \% A / H} & \textbf{Agent} & \textbf{Supervised} & \textbf{Devin} \\
\midrule
\textbf{Line, primary} & 648,959 & 28.0 / 34.5 & \textbf{0.83} [0.75, 0.93] & \textbf{0.76} [0.65, 0.87] & 0.95 [0.82, 1.11] \\
Line, with vendored paths & 854,139 & 28.1 / 34.8 & \textbf{0.83} [0.75, 0.93] & \textbf{0.76} [0.66, 0.87] & 0.95 [0.82, 1.10] \\
Line, unweighted & 648,959 & 23.4 / 30.1 & 0.91 [0.69, 1.20] & 0.80 [0.60, 1.07] & 1.06 [0.65, 1.71] \\
Line, unweighted, with vendored & 854,139 & 23.3 / 50.3 & 0.92 [0.68, 1.25] & 0.86 [0.58, 1.29] & 1.00 [0.68, 1.48] \\
Line, sampled & 277,283 & 28.1 / 34.7 & \textbf{0.83} [0.75, 0.93] & \textbf{0.76} [0.66, 0.88] & 0.95 [0.81, 1.10] \\
Hunk, primary & 60,095 & 39.5 / 44.8 & 0.93 [0.85, 1.02] & \textbf{0.82} [0.72, 0.92] & 1.11 [0.98, 1.26] \\
File, primary & 18,120 & 84.2 / 88.2 & 0.97 [0.90, 1.04] & 0.95 [0.86, 1.04] & 1.01 [0.90, 1.13] \\
\bottomrule
\end{tabular}}
\parbox{\linewidth}{\footnotesize \vspace{0.3em} Primary: vendored and generated paths excluded, PR-weighted. Sampled: primary design, at most 20 files per PR and 100 lines per file. Bold: $p<0.05$. Mod.\ \%: share of units modified by December 31, 2025 (PR-weighted in the weighted designs).}
\end{table}

\begin{figure}[t]
    \centering
    \includegraphics[width=\linewidth]{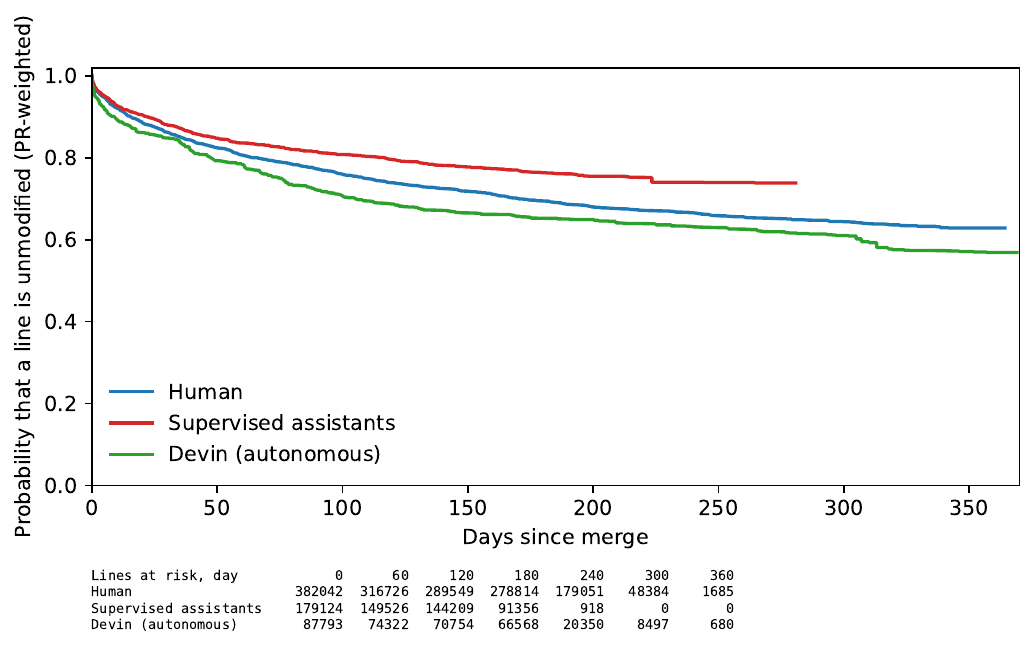}
    \caption{PR-weighted Kaplan-Meier curves of line survival (primary design). Each curve ends at the longest follow-up of its group. The earliest supervised-assistant PR was merged on March 26, 2025 (most in June and July), so no supervised-assistant line can be followed beyond day 281. The steps at day 223 (supervised assistants) and day 313 (Devin) each come from a single PR.}
    \label{fig:km_line}
\end{figure}

\begin{figure}[t]
    \centering
    \includegraphics[width=\linewidth]{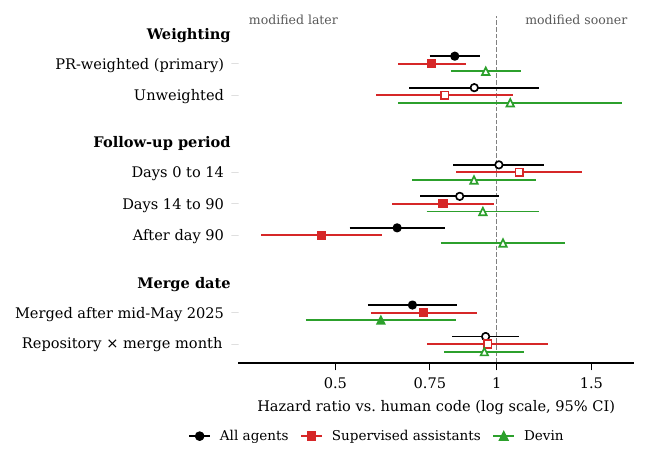}
    \caption{Within-repository Cox HR [95\% CI] of agent-attributed vs.\ human lines under the design choices of RQ1. All rows except \emph{Unweighted} are PR-weighted. Filled markers: $p < 0.05$. An HR below 1 means that agent-attributed code is modified later.}
    \label{fig:hr_forest}
\end{figure}

\paragraph{Weighting, vendored code and pooling}
Figure~\ref{fig:hr_forest} shows how the HR changes with the weighting, the follow-up period and the merge date. Including vendored paths gives the same HRs, although 34.1\% of all human lines, but only 3.8\% of supervised-assistant and 0.3\% of Devin lines, lie in such paths. The sampled design, which caps the size of a PR in another way, also gives the same HRs. However, without weighting, no difference is significant. The lower hazard of supervised-assistant code thus holds when the analysis is PR-weighted, but not when a few very large PRs dominate. A Cox model pooled across repositories, with PR-clustered standard errors, gives HR = 0.89 [0.80, 1.00] for the primary design and 0.76 [0.52, 1.10] for unweighted lines with vendored paths. So a comparison across repositories is no substitute for one within them.

\paragraph{Proportional hazards and RMST}
The Schoenfeld test rejects the proportional hazards assumption ($p < 0.001$). When we split the follow-up, the agent HR is 1.01 [0.83, 1.23] in the first 14 days, 0.85 [0.72, 1.01] from day 14 to day 90 and 0.65 [0.53, 0.80] afterwards. Agent-attributed and human lines are thus modified at the same rate in the first two weeks, and the difference builds up later. The supervised assistants show the same pattern (1.10 [0.84, 1.44], 0.79 [0.64, 0.99] and 0.47 [0.36, 0.61]), whereas Devin does not differ from human code in any period. Consistently with the early period, within 90 days agent-attributed lines stay unmodified only 0.8 days longer on average than human lines [$-2.3$, $3.8$], and 2.2 days longer within 150 days [$-3.4$, $7.9$], neither of which is significant. Within 14 days, the window of GitClear's churn~\cite{gitclearCodingCopilot}, 8.3\% of supervised-assistant lines, 12.0\% of Devin lines and 9.6\% of human lines are modified (PR-weighted).

\paragraph{Calendar time}
We acknowledge that supervised-assistant PRs were merged later (median June 18, 2025, vs.\ April 9 for human and Devin PRs), so their late follow-up falls in a different calendar period. When we keep only the PRs merged on or after May 16, 2025, the release of Codex~\cite{codexDocs}, instead of the full cohort, the HR is 0.70 [0.57, 0.84] for agent-attributed code, 0.73 [0.58, 0.92] for the supervised assistants and 0.61 [0.44, 0.84] for Devin, although only 133 Devin PRs remain. Within this period, Devin code is therefore also modified later than human code, so the difference between the tiers in the full cohort may partly reflect when their PRs were merged. However, when we stratify by repository and merge month, the supervised-assistant HR is 0.96 [0.74, 1.24]. Only 128 of the 657 strata contain both agent and human PRs, so this estimate is imprecise.

\paragraph{Hunks, files and substantial modification}
Supervised-assistant hunks are modified later than human hunks (HR = 0.82), whereas the agent group as a whole and Devin do not differ significantly (Table~\ref{tab:rq1_line}). Without weighting, Devin hunks are modified sooner (HR = 1.24 [1.00, 1.53], $p = 0.049$). At the file level, no group differs. For the substantial modification event, agent-attributed hunks and files reach substantial modification later than human ones at every $\theta$ (HR 0.80 to 0.86, all $p < 0.02$). Again, this comes from the supervised assistants (0.72 to 0.77), whereas Devin does not differ at any $\theta$ (0.93 to 1.01). Separating minor edits from rewrites therefore does not change the conclusion.

\paragraph{Attribution, file history and repository activity}
Excluding the 16 cohort PRs that the audit labeled human, or the 28 labeled human or unclear, leaves the primary estimate unchanged (0.84 and 0.83, against 0.83). The 136 audited PRs labeled agent give a similar but non-significant estimate (0.89 [0.64, 1.23]), because the sample is much smaller. Adjusting for the history of each line's file moves the agent HR to 0.88 [0.79, 0.98] and the supervised-assistant HR to 0.81 [0.70, 0.93]. About a quarter of the difference is explained by the files that the PRs change, but the difference remains. Files with more recent changes are modified sooner (HR = 1.26 per unit of $\log(1+{}$changes in the previous three months$)$), and lines in test files later (HR = 0.73). Within terciles of repository activity, the agent HR is 0.83 [0.44, 1.57], 0.63 [0.43, 0.94] and 0.86 [0.77, 0.97]. The direction is the same in every tercile, so the result does not depend on the most active repositories.

\paragraph{Task type}
The groups address somewhat different tasks (Cram\'er's V = 0.12). Fixes make up 41.3\% of the supervised-assistant PRs and 35.0\% of the human PRs, and new features 47.3\% of the Devin PRs and 39.5\% of the human PRs. Within the same repository and task type, the supervised-assistant HR is 0.77 [0.65, 0.91] and the agent HR 0.87 [0.78, 0.99], whereas Devin code still does not differ from human code (1.03 [0.87, 1.21]). Without weighting, the supervised-assistant difference is now significant (0.77 [0.61, 0.96]). Within the two largest task types, the supervised-assistant HR is 0.75 [0.57, 0.98] for fixes and 0.82 [0.63, 1.06] for features. The difference therefore does not come from the tasks that the supervised assistants are given.

\paragraph{Individual agents}
Table~\ref{tab:agent_breakdown} reports the results per agent. The lower hazard of the supervised assistants comes from Copilot (HR = 0.75) and Codex (HR = 0.69), which contribute 980 of their 1,055 PRs. Cursor and Claude Code point in the other direction, but with wide intervals, since they contribute only 44 and 31 PRs and a single PR adds 26\% and 33\% of their lines. We therefore do not interpret their estimates.

\begin{table}[t]
\centering
\caption{RQ1: Line-level results per agent (primary design). Largest PR: share of the agent's lines added by its largest PR.}
\label{tab:agent_breakdown}
\resizebox{\columnwidth}{!}{
\begin{tabular}{lrrrr}
\toprule
\textbf{Author} & \textbf{PRs} & \textbf{HR [95\% CI]} & \textbf{Modified (\%)} & \textbf{Largest PR} \\
\midrule
GitHub Copilot & 489 & \textbf{0.75} [0.59, 0.94] & 22.5 [19.6, 25.3] & 9.7\% \\
OpenAI Codex & 491 & \textbf{0.69} [0.57, 0.84] & 24.4 [21.4, 27.8] & 5.7\% \\
Cursor & 44 & 1.81 [0.91, 3.60] & 35.0 [24.6, 46.7] & 26.4\% \\
Claude Code & 31 & 1.66 [0.81, 3.41] & 30.5 [18.0, 43.5] & 32.9\% \\
Devin & 412 & 0.96 [0.82, 1.11] & 38.0 [34.4, 41.5] & 41.6\% \\
\midrule
Human & 1,327 & & 34.5 [32.6, 36.6] & 9.0\% \\
\bottomrule
\end{tabular}}
\parbox{\linewidth}{\footnotesize \vspace{0.3em} HR: within-repository Cox HR vs.\ human code. Bold: $p<0.05$. Modified: PR-weighted share of lines modified by December 31, 2025, with a 95\% PR bootstrap CI.}
\end{table}

\begin{tcolorbox}[colback=gray!10, colframe=gray!50, title=RQ1 Summary]
Agent-attributed code is not modified sooner than human code in the same repositories, in any design we examined. Lines from supervised assistants, mainly Copilot and Codex, are modified later than human lines (HR = 0.76), but only when every PR has the same weight. This difference is absent in the first two weeks and holds among PRs merged after mid-May 2025, although stratifying by repository and merge month cannot confirm it. Devin code is modified as soon as human code in the full cohort, but later among PRs merged after mid-May 2025, so the difference between the tiers may partly reflect when their PRs were merged.
\end{tcolorbox}

\section{RQ2: When Agent-Attributed Code Is Modified, What Is the Intent of the Modification?}
\label{sec:rq2}

\subsection{Motivation}
Survival alone does not show that agent-attributed code is sound. The code may stay unmodified because it works, or because its defects have not been found yet. If agent-attributed code survives only because its defects surface later, a larger share of the commits that modify it should be bug fixes. Therefore, we examine the intent of the commits that modify agent-attributed and human code.

\subsection{Approach}
\paragraph{Intent classification}
We classify each modifying commit by its message into the maintenance categories of Swanson~\cite{swanson1976dimensions}, corrective, adaptive and perfective, extended with preventive maintenance as in the maintenance standard ISO/IEC/IEEE 14764~\cite{iso14764}. We use keyword rules, the approach of Mockus and Votta~\cite{mockus2000identifying} that is common in mining software repositories research~\cite{hindle2009automatic,levin2017boosting}. Our rules first assign one of eight fine-grained categories, based on keywords from prior work on commit and change classification~\cite{barreto2024dvc,humbatova2020taxonomy,islam2019comprehensive,alomar2019can,zhou2021spi,zeng2025first}, and then map them to five classes:
\begin{itemize}[leftmargin=*]
    \item \textbf{Corrective:} bug fixes (e.g., \textit{fix}, \textit{bug}, \textit{error}, \textit{crash}, \textit{regression}).
    \item \textbf{Perfective:} refactoring (e.g., \textit{refactor}, \textit{clean}, \textit{simplify}), performance (e.g., \textit{perf}, \textit{cache}), documentation (e.g., \textit{doc}, \textit{readme}) and new functionality (e.g., \textit{feat}, \textit{add}, \textit{implement}). Our perfective class therefore includes feature additions and documentation, as in ISO/IEC/IEEE 14764, which defines perfective maintenance as providing enhancements and improvements of information for users~\cite{iso14764}.
    \item \textbf{Adaptive:} maintenance of the environment (e.g., \textit{chore}, \textit{bump}, \textit{upgrade}, \textit{dependency}, \textit{ci}, \textit{build}, \textit{config}).
    \item \textbf{Preventive:} testing and security (e.g., \textit{test}, \textit{coverage}, \textit{mock}, \textit{security}, \textit{auth}, \textit{sanitize}).
    \item \textbf{Other:} commits that match none of the keywords.
\end{itemize}
When several categories match, we choose the most specific intent with a fixed priority: security, bug fix, performance, new functionality, refactoring, testing, documentation and maintenance. For example, the message ``fix bug in config update logic'' matches both a bug fix (\textit{fix}, \textit{bug}) and maintenance (\textit{update}, \textit{config}). We classify it as corrective, because the message describes a defect, whereas \textit{config} only names the location of the fix. Keywords match whole words only. Matching substrings would, for example, let \textit{auth} match the \texttt{Co-authored-by} trailer that agents add to their commits, which would inflate the preventive class. We report substring matching as a sensitivity analysis. The complete keyword lists are in our replication package.

\paragraph{Unit of analysis}
A modifying commit often changes many added lines at once, and a single large commit can change tens of thousands of lines, which would dominate a count of lines. We therefore take one observation per modifying commit and author group. A commit that modifies agent-attributed lines counts once for the agent group, and once more for the human group if it also modifies human lines. We use the modifications of the primary analysis of RQ1, that is, of lines outside vendored and generated paths. Because we follow the first-parent history, 416 of the modifying commits are merge commits. Their messages describe the merge rather than the change, and 280 of their 456 observations fall into the adaptive class through the keyword \textit{merge}. We therefore repeat the test without merge commits.

\paragraph{Statistical analysis}
We compare the intent distributions of agent-attributed and human code with a chi-square test of independence~\cite{agresti1996introduction}, report Cramér's V as the effect size~\cite{cramer1999mathematical}, and use standardized residuals to find the classes that drive a difference~\cite{sharpe2015your}. A commit that modifies both groups appears in both columns, which violates the independence assumption of the test. Furthermore, we repeat the test on the commits that modify only one group.

\subsection{Results}

\paragraph{Intent distributions}
Table~\ref{tab:intent_distribution} reports the intent of the 6,112 observations (2,347 for agent-attributed and 3,765 for human code). The distributions differ statistically ($\chi^2 = 39.1$, $df = 4$, $p < 0.001$), but the effect size is negligible (Cram\'er's V = 0.080). Only the Other class deviates beyond $|z| > 2$: commits that modify agent-attributed code are less often unclassifiable (10.6\% vs.\ 15.9\%, $z = -4.28$). Corrective commits are slightly more frequent for agent-attributed code (35.6\% vs.\ 32.5\%), but the residual stays below the threshold ($z = +1.59$). Perfective commits are the most common class for both groups (38.3\% vs.\ 35.4\%).

\begin{table}[t]
\centering
\caption{RQ2: Intent of the commits that modify agent-attributed and human code (one observation per modifying commit and author group)}
\label{tab:intent_distribution}
\begin{tabular}{lrrrr}
\toprule
\textbf{Intent} & \textbf{Agent (\%)} & \textbf{Human (\%)} & \textbf{$\Delta$ (pp)} & \textbf{$z$} \\
\midrule
Corrective & 35.6 & 32.5 & +3.1 & +1.59 \\
Perfective & 38.3 & 35.4 & +2.9 & +1.46 \\
Adaptive & 11.7 & 12.7 & $-$1.0 & $-$0.89 \\
Preventive & 3.9 & 3.5 & +0.4 & +0.54 \\
Other & 10.6 & 15.9 & $-$5.3 & $-$4.28 \\
\bottomrule
\end{tabular}
\parbox{\linewidth}{\footnotesize \vspace{0.3em} $\chi^2 = 39.1$, $df=4$, $p<0.001$, Cram\'er's V $= 0.080$. $z$: standardized residual for agent-attributed code.}
\end{table}

\paragraph{Sensitivity analyses}
Without merge commits, the result is the same (V = 0.081), and the adaptive share drops to 8.3\% for agent-attributed and 8.4\% for human code. On the commits that modify only one group, which satisfy the independence assumption, the result is also the same (V = 0.084, Corrective 36.2\% vs.\ 32.4\%). With one observation per modified file and commit instead, the distributions do not differ ($\chi^2 = 7.0$, $p = 0.13$, V = 0.035). However, substring matching changes the result. It doubles the effect size (V = 0.155), and the preventive share becomes 29.7\% for agent-attributed and 18.8\% for human code, compared with 3.9\% and 3.5\% with whole-word matching. The reason is that \textit{auth} matches the \texttt{Co-authored-by} trailer of agent commits. In other words, the preventive difference under substring matching comes from the classifier, not from the code.

\begin{tcolorbox}[colback=gray!10, colframe=gray!50, title=RQ2 Summary]
The commits that modify agent-attributed code have almost the same intent profile as those that modify human code (Cram\'er's V = 0.080). Corrective commits are slightly more common for agent-attributed code (35.6\% vs.\ 32.5\%), but its standardized residual stays below 2. Within our observation window, we see no sign that agent-attributed code survives because its defects surface later.
\end{tcolorbox}

\section{RQ3: Can We Predict at Merge Time Whether, and When, Agent-Attributed Code Will Be Modified?}
\label{sec:rq3}

\subsection{Motivation}
If maintainers could tell at merge time which agent-attributed files will be modified, and how soon, they could, for example, ask for a second review of the files with the highest predicted risk. Such a flag is only useful if it is more precise than a simple rule. We ask whether agent-attributed files will be modified (\textbf{RQ3a}), whether the prediction points to the lines that will change, and when the modification will happen (\textbf{RQ3b}). The unit is the file, so that the model can use the content and history of the whole file at merge time. We use the files changed by agent PRs, outside vendored paths, and repeat the models on all files as a comparison.

\subsection{Approach}
\paragraph{Evaluation}
Following the advice of Tantithamthavorn et al.~\cite{tantithamthavorn2016empirical} to repeat the validation, we use 30 repetitions of 10-fold cross-validation grouped by repository, so that patterns of a repository cannot leak from the training to the test folds~\cite{pedregosa2011scikit}. We compare logistic regression, SVM, naive Bayes, KNN, random forest, XGBoost, CatBoost and an MLP with the default hyper-parameters of their libraries, and rank them with the Scott-Knott ESD test~\cite{tantithamthavorn2016empirical}. We explain the best model with SHAP~\cite{lundberg2017unified} and LIME~\cite{ribeiro2016should} out of sample: we split the repositories into five folds, train on four and explain the predictions for the fifth.

\paragraph{RQ3a}
The label of a file is 1 if any later commit on the default branch touches it before December 31, 2025. We represent the content of the file at merge time, including lines that other authors wrote, as a bag of words (at most 1,000 features). Following Rahman et al.~\cite{rahman2019natural}, we remove syntax tokens such as separators, operators and keywords, which add noise, and use unigrams to trigrams, since the cross-entropy of n-gram models of code levels off at about this order~\cite{hindle2016naturalness}. Most files are modified, so the minority class is the files that are not. We oversample it with SMOTE~\cite{chawla2002smote} in the training folds only, so that no synthetic file reaches a test fold.

As a baseline, we fit a logistic regression on the size of the file and its number of changes in the three months before the PR. To test whether the model points to the lines that will change, we score each added line by the sum of the out-of-sample SHAP values of its tokens, as in explainable defect prediction~\cite{pornprasit2021pyexplainer,tantithamthavorn2021explainable}, and compare this ranking with ranking lines by their length, since a longer line has more tokens that can change.

\paragraph{RQ3b}
\label{sec:rq3b}
For the modified files, we classify the time to modification into four bins: within one day, one day to one week, one week to one month, and more than one month. These bins are similar to those of Hasan et al.~\cite{hasan2023understanding}. We chose them because they match how maintainers plan work, from the same day to the next release, and because no bin is rare in our data (the smallest bin holds 15\% of the files). We report the Macro F1, so that each bin counts equally, and the macro AUC-ROC. The features are the process features of Khatoonabadi et al.~\cite{khatoonabadi2024predicting}, measured at the merge time $T$: the commits of the project and the changes to the file in the previous three months, the age of the file, the number of open PRs in the repository at $T$, the earlier PRs of the PR author in the repository (created before $T$, excluding the PR itself) and the share of them merged before $T$, and the day and hour of the merge. We drop the files whose author history could not be retrieved (22 of the 6,300 modified agent files, 0.3\%), and remove correlated features with AutoSpearman~\cite{jiarpakdee2018autospearman}.

\subsection{Results}

\paragraph{RQ3a: file-level prediction}
After removing comments and syntax tokens, and the files in unsupported languages or left without tokens, the agent dataset contains 6,661 files from 155 repositories, of which 79.7\% are modified. Random forest, CatBoost and XGBoost form the first Scott-Knott rank, with an AUC-ROC of about 0.70 and an AUC-PR of about 0.89, and the other five models reach an AUC-ROC of 0.51 to 0.66. The AUC-PR looks high mainly because about 80\% of the files are modified anyway, which is the AUC-PR of a random classifier. However, the logistic regression baseline on the size of the file and its changes in the previous three months performs clearly better (AUC-ROC 0.78, AUC-PR 0.93). In other words, the tokens of the code add nothing beyond what the size and the recent history of a file already tell us. The same holds for all 15,946 files (83.8\% modified): random forest, XGBoost and CatBoost reach an AUC-ROC of about 0.69, and the baseline reaches 0.75. The tokens with the largest SHAP values, such as \texttt{key}, \texttt{model}, \texttt{config}, \texttt{error} and \texttt{state}, point to configuration and integration code, and LIME agrees only partly (46.9\% of its top tokens are among the 50 with the highest SHAP value). A LIME explanation takes about 0.2 seconds per file.

\paragraph{RQ3a: line localization}
Of the agent files, 1,941 files from 109 repositories have both modified and unmodified added lines (163,409 lines, 22.8\% modified). The ranking of lines by their SHAP scores barely beats chance (AUC-ROC 0.546, per-file mean 0.576), and ranking lines by their length does as well (0.552). Among the top 10\% of the lines of each file, 36.7\% are later modified, compared with 30.1\% for a random choice and 39.7\% for the longest lines. The random value exceeds the overall share of modified lines because it averages over files, and small files have a higher share.

\paragraph{RQ3b: time to modification}
The agent dataset contains 6,278 modified files from 160 repositories, of which 15.1\% are modified within one day, 25.0\% within one week, 26.8\% within one month and 33.1\% later. AutoSpearman keeps all eight features. XGBoost, CatBoost and random forest form the first Scott-Knott rank, with a Macro F1 of about 0.31, against 0.25 for a random guess and 0.27 to 0.28 for the other models, and a macro AUC-ROC of about 0.60 (in the five-fold split used for the explanations, XGBoost reaches 0.32 and 0.61). For all files, CatBoost alone ranks first (Macro F1 0.32, AUC-ROC 0.62). SHAP ranks the changes to the file in the previous three months first, followed by the number of open PRs at merge time and the age of the file, and LIME agrees on the leading feature. The history of the PR author ranks fifth and sixth of the eight features.

\begin{tcolorbox}[colback=gray!10, colframe=gray!50, title=RQ3 Summary]
Code tokens predict whether an agent-attributed file will be modified with modest accuracy (AUC-ROC 0.70), but a baseline on the file's size and recent changes predicts it better (AUC-ROC 0.78). The file-level model does not localize the lines that are later modified, since ranking lines by their length does as well. Process features predict when agent-attributed code will be modified only weakly (Macro F1 0.31 against 0.25 for chance), and the recent changes to the file are the strongest predictor.
\end{tcolorbox}

\section{Discussion}
\label{sec:discussion}

\subsection{Is Agent-Attributed Code Disposable?}
Within the same repositories, agent-attributed code is not modified sooner than human code in any design we examined. Therefore, the disposable code hypothesis is not supported for the agents and repositories in our cohort. However, the opposite claim, that agent-attributed code survives longer, holds only in part. Firstly, it holds in the PR-weighted analysis, but not when a few large PRs dominate. Secondly, it comes from Copilot and Codex, whereas Devin code is modified as soon as human code in the full cohort. Thirdly, it is absent in the first two weeks and largest after 90 days, and about a quarter of it is explained by the files that the PRs change. Survival also does not imply quality, since stable code may lie in rarely changed parts of the system. However, RQ2 shows that agent-attributed code is modified for much the same reasons as human code, so we see no sign that its defects surface later within our observation window.

\subsection{Supervised Assistants and Devin}
The lower hazard of the supervised assistants comes from Copilot and Codex, while Cursor and Claude Code contribute too few PRs to tell. In AIDev, Copilot's PRs are also opened by its own account (\texttt{Copilot}), so the bot identity cannot explain the difference between Copilot and Devin. We see two explanations. Firstly, the tiers were used in different periods, and among PRs merged after mid-May 2025 Devin code is also modified later than human code (Section~\ref{sec:rq1}). Secondly, we conjecture that the difference lies in who takes responsibility for the code. A Copilot or Codex task is started by a developer of the repository, who then reviews the result, whereas Devin is offered primarily as an engineer that takes tasks from a team. Prior work found that components with clear owners have fewer defects~\cite{bird2011don,greiler2015code}, and code that a developer feels responsible for may also need fewer follow-up changes. We did not measure ownership, so this explanation remains a hypothesis. The kind of code may also matter: 35.6\% of supervised-assistant lines are in test files, compared with 24.1\% of human lines, and test lines are modified later. However, adjusting for test paths and file history does not remove the difference.

\subsection{Why Do Our Results Differ from Concurrent Studies?}
\label{sec:disc_disagree}
Our conclusion agrees with Kraishan~\cite{kraishan2026agents}, who found that no agent churns more than humans, and with the null overall result of Xia and Miller~\cite{xia2026these}. However, other studies found more corrective maintenance of agentic code~\cite{xia2026these}, fewer fully surviving agent commits~\cite{popescu2026investigating} and more follow-up fix PRs~\cite{takerngsaksiri2026finishes}. These studies differ from ours in the unit, the event, the window, the period and the intent classifier. For example, Popescu et al. follow the first commit of each PR for three weeks, the period in which we also find no lower hazard of agent-attributed code. These choices may explain part of the differences, which we did not test.

\subsection{Forecasting}
The tokens of agent-attributed code say something about whether a file will be modified, but less than its size and recent history do (Section~\ref{sec:rq3}). The model also does not point to the lines that will change, since ranking lines by their length does as well. This matters because localizing lines from explained file-level models is established in defect prediction~\cite{pornprasit2021pyexplainer}, but for code modification it does not transfer when evaluated out of sample. When a file will be modified is even harder to predict, and the best predictor is again the recent history of the file, not who wrote it. Consistent with prior work~\cite{graves2002predicting}, a flag at merge time is better built on the history of a file than on its content or its author.

\subsection{Implications}
\textbf{For practitioners.} Firstly, code merged from Copilot and Codex in our cohort is modified no sooner than human code in the same repositories, so teams need not treat it as throwaway code. Secondly, Devin code is modified somewhat more often within two weeks (12.0\% vs.\ 9.6\% of lines), which may reflect when its PRs were merged.

\textbf{For researchers.} Firstly, agents differ, and grouping them as ``AI'' hides these differences, so studies should report results per agent or per way of working with the agent. Secondly, survival comparisons should compare code within repositories and report how they weight units, since the weighting can change the conclusion. The same holds for intent, where substring matching creates a spurious difference.

\section{Threats to Validity}
\label{sec:threats}

\textbf{Construct validity.}
We acknowledge that AIDev attributes a PR to an agent from PR-level signals, and that a developer may have written part of its code. Our audit addresses this threat: the agent made substantive changes in 87.3\% of the audited PRs in our cohort, and removing the PRs labeled human, or human and unclear, leaves the RQ1 estimate unchanged (Section~\ref{sec:rq1}). A second rater labeled 100 of these PRs, with almost perfect agreement (Cohen's $\kappa$ = 0.85). Conversely, human PRs may contain undisclosed AI-assisted code~\cite{xiao2026self}, so we compare agent-attributed code with code that is not attributed to an agent.

We count any change to a line as a modification, including trivial edits, which the substantial modification event addresses for hunks and files. Furthermore, we run \texttt{git blame --reverse} without move and copy detection. It follows the lines of a renamed file, but it counts a moved block of lines as modified, and at the file level a renamed file counts as modified. We expect this to affect both groups alike, but we did not verify it. Finally, keyword classification of commit messages has moderate accuracy in prior work~\cite{levin2017boosting}, and we did not validate our classifier against manual labels. Therefore, RQ2 compares coarse intent profiles, and its corrective shares should not be read as defect rates.

\textbf{Internal validity.}
We acknowledge that agent and human PRs may address different tasks within the same repository. We stratify by repository and adjust for PR size, and in sensitivity analyses we adjust for file history and stratify by the task type that AIDev assigns to each PR, which leaves the supervised-assistant difference unchanged. However, these labels come from the PR title and description, and other differences between tasks, such as their difficulty, remain uncontrolled. We could not mine 359 merged PRs and two repositories (Section~\ref{sec:cohort}). Our cohort also contains merged PRs only, so our results describe the code that passed review, not all the code that agents write. Furthermore, supervised-assistant PRs were merged later than human and Devin PRs (median June 18 vs.\ April 9, 2025), so their late follow-up falls in a different calendar period. Among PRs merged after mid-May 2025, the supervised-assistant difference remains and Devin code is also modified later, but stratifying by repository and merge month cannot confirm it (Section~\ref{sec:rq1}). In RQ3, the features describe the whole file, including lines of other authors, and the models use default hyper-parameters, so tuned models may do somewhat better.

\textbf{Conclusion validity.}
Our many sensitivity analyses increase the chance of a spurious significant result. We therefore base our conclusions on the primary analysis and on the direction of the estimates across variants. Cursor and Claude Code contribute fewer than 50 PRs each, so their estimates are imprecise.

\textbf{External validity.}
Our findings may not generalize beyond our 215 repositories, for example to small or closed-source projects, or to agents released after August 2025.

\section{Related Work}
\label{sec:related}

\paragraph{AI-generated code and coding agents}
Research on AI-generated code has mostly evaluated it at generation time, with metrics such as Pass@k~\cite{chen2021evaluating} and benchmarks such as SWE-bench~\cite{jimenez2023swe}. Other studies found vulnerabilities and maintainability issues in generated code~\cite{pearce2025asleep,sandoval2023lost,yeticstiren2023evaluating}. The AIDev dataset~\cite{li2026aidev} made it possible to study the PRs that coding agents open on GitHub. These studies concern code before or at the merge, whereas we follow it after the merge.

\paragraph{Post-merge fate of agent code}
Popescu et al.~\cite{popescu2026investigating} measured how many lines of the first commits of agent and human PRs survive, change or are deleted after three days, one week and three weeks. They found that human commits more often survive in full, with small effect sizes. Kraishan~\cite{kraishan2026agents} measured churn as the follow-up commits on the most-changed file of a PR within 90 days and found that no agent churns more than humans, although Devin PRs are reverted more often. Sawada et al.~\cite{sawada2026extent} found that AI-generated files are most often modified to extend features, mostly by human developers. Xia and Miller~\cite{xia2026these} followed added lines in 182 repositories with repository-stratified Cox models. They found that agentic code is terminated at a rate similar to human code (HR = 1.11, not significant), but receives more corrective maintenance. Takerngsaksiri et al.~\cite{takerngsaksiri2026finishes} found that agent PRs are more often followed by a fix PR within 30 days (3.68\% vs.\ 2.34\%).

Unlike these studies, we compare every added line with human lines in the same repository, give each PR the same weight, report each agent separately and ask whether the fate of agent-attributed code can be predicted at merge time.

\paragraph{Survival analysis and time-to-event prediction in software engineering}
Survival analysis has been used to estimate the lifetime of lines and tokens of code~\cite{spinellis2021software}, to prioritize static analysis warnings~\cite{aman2019survival} and to model how long open-source projects stay active~\cite{samoladas2010survival}, and related studies track how long vulnerabilities persist~\cite{dipenta2009life}. When an event will happen has been predicted for bug fixes~\cite{zhang2013predicting}, developer retention~\cite{lin2017developer} and the first response to a PR~\cite{khatoonabadi2024predicting}, whose process features we adapt in RQ3.

\section{Conclusion}
\label{sec:conclusion}
In this paper, we followed 648,959 lines added by 2,794 merged PRs in 215 repositories from AIDev to test whether agent-attributed code is modified sooner than human code in the same repositories. It is not, in any design we examined. In the PR-weighted analysis, lines from supervised assistants, mainly Copilot and Codex, are modified later than human lines, whereas Devin code does not differ in the full cohort but is also modified later among PRs merged after mid-May 2025. When agent-attributed code is modified, the intent is much the same as for human code. The size and recent history of a file predict whether agent-attributed code will be modified better than its code does, but neither which of its lines will change nor when it will be modified can be predicted well at merge time. Finally, the lower hazard of supervised-assistant code depends on weighting PRs equally, so studies of code survival should report how they weight the code they compare.

In the future, we plan to test the ownership explanation, for example by surveying developers, and to replicate the analysis in closed-source projects.

\section*{Data Availability}
\label{sec:replication}
Our data, attribution audit and analysis scripts are available at \url{https://anonymous.4open.science/r/will-it-survive-replication-AB78/}.

\section*{Acknowledgement}
We used AI-enabled tools (Claude and Grammarly) to assist with manuscript revisions. All AI-assisted content was reviewed and verified for accuracy by the authors.

\bibliographystyle{ACM-Reference-Format}
\bibliography{references}

\end{document}